\documentclass{sigchi}

\usepackage{textcomp}
 \toappear{Permission to make digital or hard copies of all or part of this work for personal or classroom use is granted without fee provided that copies are not made or distributed for profit or commercial advantage and that copies bear this notice and the full citation on the first page. Copyrights for components of this work owned by others than ACM must be honored. Abstracting with credit is permitted. To copy otherwise, or republish, to post on servers or to redistribute to lists, requires prior specific permission and/or a fee. Request permissions from Permissions@acm.org.\\
 {\emph{UbiComp '15}}, September 7--11, 2015, Osaka, Japan. \\
Copyright 2015 \textcopyright ACM 978-1-4503-3574-4/15/09...\$15.00.  \\
http://dx.doi.org/10.1145/2750858.2804252.}

\usepackage{graphicx}
 
\toappear{arXiv pre-print. To appear in \emph{MobileHCI '17}, September 04--07, 2017, Vienna, Austria} 

\def\plaintitle{Productive, Anxious, Lonely -\\24 Hours Without Push Notifications}
\def\plainauthor{Martin Pielot and Luz Rello}
\def\plainkeywords{Notifications; Mobile Devices; Deprivation Study}
\def\plaingeneralterms{Human Factors}

\newcommand{\compresslist}{
  \setlength{\itemsep}{3pt}
  \setlength{\parskip}{0pt}
  \setlength{\parsep}{0pt}
}

\usepackage[usenames,dvipsnames]{xcolor} 
\usepackage{color}
\definecolor{Orange}{rgb}{0.9,0.5,0}

\newcommand{\iw}{1} 

\newcommand{\statement}[1]{``\emph{#1}''} 
\newcommand{\finding}[1]{#1} 
\newcommand{\result}[1]{(#1)} 
\newcommand{\para}[1]{\textbf{#1:}} 

\usepackage{balance}  
\usepackage{graphics} 
\usepackage[T1]{fontenc}
\usepackage{txfonts}
\usepackage{times}    
\usepackage[pdftex]{hyperref}
\usepackage{url}      
\usepackage{color}
\usepackage{textcomp}
\usepackage{booktabs}
\usepackage{ccicons}
\usepackage{todonotes}

\makeatletter
\def\url@leostyle{%
  \@ifundefined{selectfont}{\def\UrlFont{\sf}}{\def\UrlFont{\small\bf\ttfamily}}}
\makeatother
\def\pprw{8.5in}
\def\pprh{11in}

\definecolor{linkColor}{RGB}{6,125,233}
\hypersetup{%
  pdftitle={\plaintitle},
  pdfauthor={\plainauthor},  
  pdfkeywords={\plainkeywords},
  pdfsubject={\plaingeneralterms},
  bookmarksnumbered,
  pdfstartview={FitH},
  colorlinks,
  citecolor=black,
  filecolor=black,
  linkcolor=black,
  urlcolor=linkColor,
  breaklinks=true,
}

\begin{document}

\title{\plaintitle}

\numberofauthors{2}
\author{
  \alignauthor{
  	\textbf{Martin Pielot}\\
  	\affaddr{Telefonica Research}\\
  	\affaddr{Placa Ernest Lluch i Marti, 5}\\
  	\affaddr{Barcelona, 08019, Spain}\\
  	\email{martin.pielot@telefonica.com}
  }\alignauthor{
  	\textbf{Luz Rello}\\
  	\affaddr{Human-Computer Interaction Institute}\\
  	\affaddr{Carnegie Mellon University}\
  	\affaddr{5000 Forbes Avenue}\\ 
  	\affaddr{Pittsburgh, PA 15213, USA}\\
  	\email{luzrello@cs.cmu.edu}
  }
}

\maketitle

\begin{abstract}

We report from the \emph{Do Not Disturb} Challenge where 30 volunteers disabled notification alerts for 24 hours across all devices. 
The effect of the absence of notifications on the participants was isolated through an experimental study design: we compared self-reported feedback from the day without notifications against a baseline day. 
The evidence indicates that notifications have locked us in a dilemma:
without notifications, participants felt less distracted and more productive.
But, they also felt no longer able to be as responsive as expected, which made some participants anxious. 
And, they felt less connected with one's social group.
In contrast to previous reports, about two third of the participants expressed the intention to change how they manage notifications. Two years later, half of the participants are still following through with their plans.
\end{abstract}

. 

\section{Introduction}

In 2010, Iqbal and Horvitz \cite{Iqbal2010} published a report of a study where they asked 20 of employees of a large IT organization to disable notifications of their work email client. While some participants realized that without notifications, they could better focus and interrupted their primary tasks less often, every one of them reenabled notifications after the study.

In 2017, notifications are no longer confined to email at work or SMS on mobile phones. They have become ubiquitous and essential to an increasing number of services, applications, and devices. People deal with dozens of notifications per day and typically attend to them within minutes \cite{Battestini2010,Pielot2014attpred,Sahami2014}, which means that they routinely interrupt concurrent activities. 

Such interruptions have shown to have negative effects on task performance in the work context \cite{Adamczyk2004,Borst2015,Czerwinski2004,Iqbal2008,Iqbal2010,Mark2012,Rennecker2005,Stothart2015}. Most notably, Iqbal and Bailey \cite{Iqbal2010}, Mark \emph{et al.} \cite{Mark2012}, and Kushlev \emph{et al.} \cite{Kushlev2016} found that notifications have negative effects on well-being as well.

In the context of mobile phones, notifications have been studied with focus on mobile messaging/SMS \cite{Birnholtz2012,Church2013,Patterson2008} and mobile phone notifications in general \cite{Pielot2014notif,Pielot2015dnd,Sahami2014}. 
However, with the notable exception of Kushlev \emph{et al.} \cite{Kushlev2016}, who asked participants to silence their phones and keep them out of sight, existing studies are limited to observations: while they can establish correlations between notifications and other factors, they cannot isolate notifications as cause. 
For example, even though Pielot \emph{et al.} \cite{Pielot2014notif} found that receiving more email notifications correlates with higher levels of stress, the study cannot isolate notifications as causal factor. Increases in stress and number of notification could have both been subject to \emph{e.g.} higher demands at work.

To better understand the effects of notifications in a holistic setting, we launched the \emph{Do Not Disturb} Challenge.
We asked 30 people to disable notifications across all sources of notifications for one day. Data was collected via questionnaire and a post-hoc interview. To isolate notifications as cause, we designed the \emph{Do Not Disturb} Challenge as an experiment and compared survey results to a baseline day. This allows us to attribute significant differences in the survey responses to the presence or absence of notifications.

The main contributions of this work are: 
\begin{itemize} \compresslist
	\item evidence that the absence of notifications has positive effects, such as making people feel less distracted and more productive;
	\item evidence that the absence of notifications also has negative effects, as people feel less connected with others and become anxious to no longer be able to adhere to social norms regarding responsiveness; and
	\item in contrast to previous work -- 73.3\% of the participants expressed the intention of disabling some notifications. Two years later, half of the participants are still following through with these plans.
\end{itemize}

\section{Related Work}

\begin{figure}[th]
\begin{center}
	\includegraphics[width=\iw\linewidth]{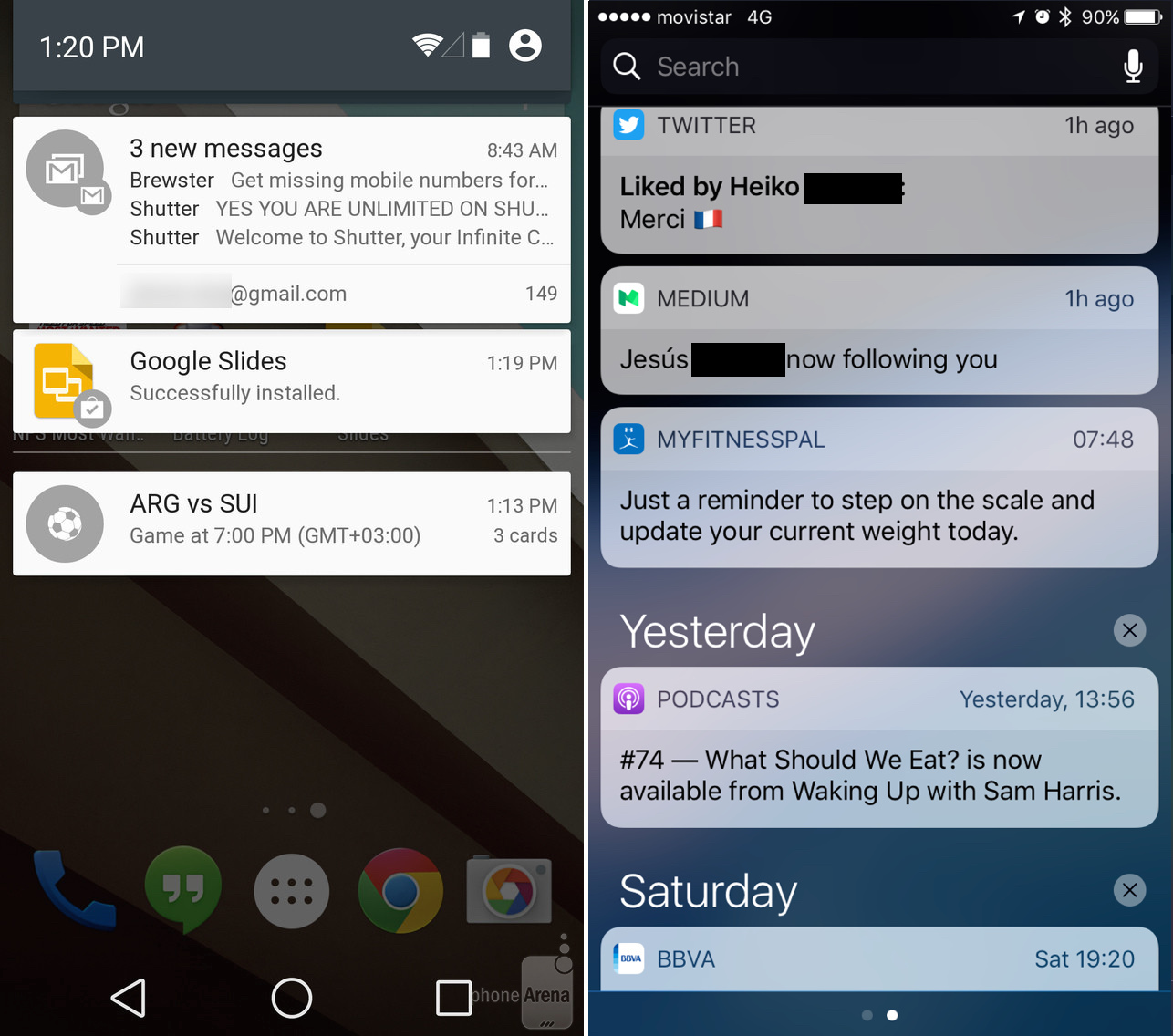}
	\caption{Notifications on Android and iOS.}
	\label{fig:notifications}
\end{center}
\end{figure}

Iqbal and Bailey \cite{Iqbal2010a} define \emph{notification} as a visual, auditory, or tactile \textbf{alert} designed to attract attention.
In daily language, the word notification may be used to describe the alert as well as a visual representation that is typically found in a pop-up or a notification center (see Figure \ref{fig:notifications}).
In this paper, we will use the word \emph{notification} to \textbf{refer to the actual alert}.

\para{Notification-Management Strategies}
In a recent survey \cite{Gallud2015}, the majority of respondents considered themselves to typically receive 20-50 or 50-100 notifications per day. 
In an \emph{in-situ} log study on mobile phone notifications \cite{Pielot2014notif}, participants received a medium number of 63.5 of notifications per day.
Both results reveal that, on average, people deal with dozens of notification alerts every day.
%
To manage this volume of notifications, Chang and Tang \cite{Chang2015} found that the ringer mode is a frequently-used mechanism to manage attentiveness to notifications on mobile phones. 
Lopez-Tovar \emph{et al. }\cite{Tovar2015} argue that users desire more fine-grained control over how notifications are presented in different contexts.  
However, Westermann \emph{et al.} \cite{Westermann2015} report that only a small fraction (10\%) of people use more sophisticated settings, such as changing notification settings for individual apps.
Thus, people typically remain exposed to the majority of the notifications alerts that they receive.

\para{Notifications \& Engagement}
Lee et al. \cite{Lee2014} showed that notifications often trigger engagement with the mobile phone: in their data set, the majority (79\%) of sessions were preceded by notifications.
As shown by Mark \emph{et al.} \cite{Mark2012}, when information workers are without (the interruption of) emails, they switch less between tasks. 
Similarly, Iqbal and Horvitz \cite{Iqbal2010} found that disabling email notifications leads to less frequent opportunistic email checking.  
Yet, not all interruptions are notification-triggered: in 2009, Jin and Dabbish \cite{Jin2009} found that in the case of information workers, 50\% of all interruptions are self-initiated. 
In addition, Oulasvirta \emph{et al.} \cite{Oulasvirta2012} report that people frequently check their phones even if there are no notifications. 
We hypothesize that disabling notifications will reduce the engagement with the mobile phone, but not eliminate it.

\para{Distraction \& Productivity Impairment}
Since people receive plenty of notifications, by sheer probability, notifications are bound to appear from time to time while the receiver is busy with other tasks. Since people usually attend to notifications within minutes, notifications may sometimes interrupt those other tasks. Such interruptions can have negative effects: previous work in the context of information workers has found that notifications negatively affect work efficiency when delivered in the middle of a work task \cite{Adamczyk2004,Czerwinski2004,Iqbal2008,Iqbal2010,Leiva2012,Rennecker2005,Stothart2015}, and the effect is more pronounced when the task is cognitively demanding \cite{Cutrell2001,Mark2015}. 
As found by Stothart \emph{et al.} \cite{Stothart2015}, this is even true when the notification is not attended, as tested in a controlled exam setting.
Hence, previous work consistently highlights the disruptive effects of notifications in work settings. 
Mobile phone users also expressed to frequently feel interrupted by notifications, even outside of work settings \cite{Pielot2014notif}.
We hypothesize that the absence of notifications will have positive effects on productivity.

\para{Notifications \& Stress}
In addition to negative effects on work efficiency, interruptions can also affect people emotionally.
Interruptions in the workplace have been linked to frustration \cite{Iqbal2008} and stress \cite{Mark2008}.
In the context of notifications, information workers felt significantly less stressed without email \cite{Mark2012}, without email notifications \cite{Iqbal2010}, or when checking work email was restricted to 3 times per day \cite{Kushlev2014}. 
However, work emails no longer reach us at only work. Our mobile devices may notify us about incoming emails at any time, which blurs the boundaries between work and private life \cite{Cecchinato2015,Chen2014}. Stress levels were found to positively correlate with the number of mobile phone notifications from, in particular, email clients \cite{Pielot2014notif}, which indicates that email notifications are particularly problematic. Mobile phone notifications in general have been linked to inattention and hyperactivity \cite{Kushlev2016}. On the basis of this related work, we hypothesize that the absence of notifications will reduce stress and other negative emotions.

\para{Notifications \& Availability}
On mobile phones, the largest chunk of notifications originate from messaging applications \cite{Lee2014,Pielot2014notif,Sahami2014}, such as SMS, WhatsApp, or Facebook Messenger.
On such communication channels, \statement{people are assumed to be constantly co-present, and thus, constantly available for conversation}  \cite{Birnholtz2012}.
On average, notifications from messengers are attended within minutes \cite{Battestini2010,Dingler2015,Gallud2015,Pielot2014attpred,Sahami2014}, and people maintain this levels of attentiveness for large parts of their wake time \cite{Dingler2015}.
Consequently, notification-enabled computed mediated communication plays a \statement{crucial role [..] in the fragmentation of the working day} \cite{Wajcman2011}.
We hypothesize that the absence of notifications will affect the participants' ability to maintain the usual level of attentiveness.


\para{Suppressing Notifications}
Two previous studies applied the methodology of depriving participants from notifications:
Iqbal and Horvitz \cite{Iqbal2010} asked 20 information workers to turn off email notifications on their work computers for one week. Compared to a baseline week, some participants checked emails more frequently as a result, but for the majority of the participants, it reduced the frequency of opportunistic email checking.
While the participants were aware that notifications are disruptive, they valued the awareness they provide. After the study, none of the participants kept notifications disabled.

Kushlev \emph{et al.} \cite{Kushlev2016} conducted a study in which for one week, 221 
participants were asked to maximize interruptions through their phone (enabling alerts, keeping phone in reach) and compared this to a baseline condition, where the same participants were asked to minimize interruptions (disabling alerts, keeping phone out of sight). The results show that with maximized interruptions, participants reported higher levels of inattention and hyperactivity -- symptoms associated with ADHD.

We complement these previous works by presented evidence that does not only focus on a single domain (work) or a single device (mobile device), but concerns the effect of notifications across different domains and devices. 
In contrast to findings by Iqbal and Horvitz, our study goes beyond work email, and finds that -- in particular related to their mobile phones -- participants formed the intention to reduce their exposure to notifications -- temporarily and permanently.
In contrast to Kushlev \emph{et al.}, our sample represents information workers from different countries instead of university students, and our combination quantitative and qualitative analysis allowed us to better understand the \emph{why} behind the quantitative findings. Neither of those previous studies touch upon the topic of maintaining availability in the context of computer-mediated communication.

\section{Methodology}

To create an experimental study about the effect of notifications across devices, we asked people to join a \emph{notification detox}: 
disable notifications for a day across all devices (experimental condition), and compare this day to a normal baseline day (control condition). Inspired by the \emph{Do Not Disturb} mode of iOS and OS X, we called the study the \emph{Do Not Disturb} Challenge. The study took place in the first half of 2015.

\begin{figure}[th]
\begin{center}
	\includegraphics[width=0.7\linewidth]{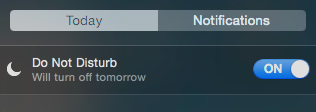}
	\caption{Do Not Disturb mode in OS X. When turned on, notification alerts are suppressed.}
	\label{fig:DnD}
\end{center}
\end{figure}

\subsection{Design}
The presence of notifications served as independent variable with two levels: in the control condition, notifications settings were left unaltered. In the experimental condition, notifications were disabled across all computing devices, applications, and services. 
The experiment used a repeated-measures design: each participant contributed to both conditions.
We counter-balanced the order in which the participants went through those two conditions to cancel out sequence effects. That is, half of the participants started the study in the experimental condition, the other half in the control condition.

Initially, we had intended to run each condition for one week. However, when we started the recruitment, many people declined participation, because they did not want to be without notifications for a whole week. To avoid a potential self-selection bias, we limited the duration of each condition to 24 hours.

Unlike Kushlev \emph{et al.} \cite{Kushlev2016}, we did not ask participants to maximize exposure to notifications in the control condition. Using the usual behavior as baseline better reflects current practices and improves the ecologic validity of our findings.

\subsection{Study Questionnaires}
We used questionnaires to collect data on the participants reaction to both conditions.
Table \ref{fig:ResultsTable} shows the statements from both questionnaires. Participants rated their level of agreement to each statement on a 5-point Likert-scale,  ranging from \emph{disagree} (score: 1) and \emph{agree} (score: 5). They were grouped along the following five aspects:
\begin{itemize} \compresslist
	\item responsiveness -- as we hypothesized that notifications are essential to maintain responsiveness,
	\item productiveness and distraction -- as previous work had revealed that the absence of notifications lead to more focus and time on task,
	\item missing information and anxiety -- as we were curious to what extent people would (worry to) miss important information and how this would affect them,
	\item stress - as previous work linked notifications to stress,
	\item social connectedness - as previous work revealed that most notifications originate from communication applications.
\end{itemize}

To not bias participants towards a positive or negative attitude while responding to the questionnaires, we balanced the number of positive and negative statements.
For example, the statement \emph{I felt distracted} was counter-balanced with \emph{I felt productive}.
The order of statements was automatically randomized to avoid sequence effects in the responses.

We opted for the use of single item measures as opposed to full questionnaires in order to keep the burden on the participants reasonable. 
The rationale for this decision was our understanding that single item measures are useful when the construct is unambiguous \cite{Wanous1997} or when a holistic impression is sufficiently informative \cite{Youngblut1993}.

\subsection{Post-Hoc Interview}
The post-hoc interview aimed at ensuring that the participants had followed the instructions for the respective study condition, collecting in-depth explanations of the questionnaire responses, and uncovering important themes that we had not considered in the questionnaires.

To this end, we conducted a semi-structured interview which was structured along the following open questions:
\begin{itemize} 
	\item How do you deal with notifications in general?
	\item What were your expectations towards the study?
	\item How was the experience to be with(out) notifications?
	\item Did you tamper with the notification settings?
	\item Is disabling notifications something you would do more often in the future? Why (not)?
\end{itemize}

With respect to Question \#4, we strongly emphasized that not complying with the rules of the respective condition would have no negative consequences for the participants, and that from a scientific point of view it was essential to know the truth. We therefore assume that participants reported truthfully.

The interviews were audio-recorded. We used Thematic Analysis \cite{Braun2006} to identify the most prominent patterns and themes within the interview data. We report them alongside the quantitative results where applicable.

\subsection{Ethics}
We recognized the possibility that participants might miss urgent and important information -- with potentially severe consequences -- which they would not have missed with notifications enabled. 
Hence, we strongly emphasized this as a risk in the informed consent.
Further, we showed participants how to set up the phone so that phone calls of selected people would still be received, in case they expected important calls. 
None of the participants made use of this option.

\subsection{Procedure}
The study took place during the first half of 2015.
Before taking part in the study, we sent the consent form to those who were interested in joining the study.
Once those people had read the consent form, and if they agreed to take part in the study, we assigned each them a participation ID to decouple their identity from their responses.
As first step, participants filled out a pre-study questionnaire, which, \emph{e.g.}, collected demographic information.

We then walked participants through all their devices and applications that create notifications and made sure that they knew how to disable them.
To ensure the absence of alerts, we required the following steps:
(1) Computing devices running iOS or OS X were set into \emph{Do Not Disturb} Mode,
(2) Android devices (only OS 5.0 and newer) were set into \emph{Priority} Mode, and
(3) finally we assisted the participants in finding settings or strategies for disabling notifications in applications that were not affected by above settings, such as Outlook or Skype.
These steps ensured that on arrival of a new notification, including phone calls, there were no auditory, haptic, or visual alerts. 

Together with the participants, we identified two consecutive study days. These days had to be working days, where no extraordinary events would take place.
We instructed participants to set up their devices in the late evening prior to each study day, so that they would start the new day within the given condition.
Depending on the condition, they would disable notifications or leave them as usual.
After 24 hours, in the late evening again, the first questionnaire was filled out and participants switched conditions.
After another 24 hours, the second questionnaire was filled out and the experimental part of the study was over.
Finally, we invited people for an open post-study interview. We let another 24 hours pass before conducting the interview to allow people to reflect on both conditions. 

\vfill\eject
\subsection{Participants}
The participants were recruited from social networks and by using the snow-ball principle. 
30 people (14 female, 16 male) volunteered to take part in the study. Their ages ranged from 19 to 56 ($M=28.9$, $SD=7.13$). 12 participants had office jobs (\emph{e.g.}, marketing manager, data analyst, ...), 8 were students, 5 were university faculty members, and 5 were working in the medical field. 
Hence, the participant sample represents white-collar workers.
With a sample size of 30 subjects, the study achieves a power of 83.2\% for detecting medium effects and 99.4\% for large effects.
None of the participants had a special motivation for participating in the study, such as a prior desire to do a notification detox.

\subsubsection{Usual Behavior}

In the pre-study questionnaire, most participants (25) reported to be {able to check notifications in most situations}, including at work. Even if not required at work, most participants still checked their mobile phones regularly in the work place. 
Thus, most participants were potentially exposed to notifications at any time of the day.

In the interview, 10 of the 30 participants reported to manage notifications of their mobile phones consciously through the ringer mode, such as
\statement{I always keep my phone in vibration mode at work} (P07), 
\statement{Normally the phone is mute, but the LED lights up when there is a notification} (P15), or 
\statement{3-4 years ago, I decided to always keep the phone in silent mode} (P25).
Rare (n=3) forms of management included disabling notifications, such as 
\statement{I usually have DnD at work and phone in silent otherwise } (P30), 
\statement{I only have notifications for Line, WhatsApp, Messenger, Calendar, and alarms} (P32), or
\statement{I only have notifications on the iPad, on the phone and the PC, they are off, except email on the PC} (P34).
The majority of the participants did not report the use of any conscious notification-management strategy. None of the participants mentioned any notification-management strategy related to stationary computers or browsers.

\subsubsection{Expectations Towards the Study}

The participants' expectations towards the study varied strongly.
15 of the 30 participants agreed with the statement of being afraid to miss urgent or important information during the day without notifications; the other half disagreed with said statement.
For example, P09 stated that \statement{I am afraid to be considered `rude' if I do not reply timely.} 
In contrast, P03 was not anxious, saying that \statement{I think I am an outlier: not many people expect fast responses; if they do, they call}.
10 participants named the boss as source of concerns: 
P10 said that since \statement{My boss was not here, so [participating] was fine.}

6 participants informed their superiors and asked for permission to take part in the study, since \statement{Notifications from my boss need to be replied to immediately} (P07). 
8 participants informed peers of taking part in the \emph{Do Not Disturb} Challenge. For example, P10 informed his girlfriend that he \emph{``probably won't respond as fast as usual.''}  and P08 reported that \emph{``I had a lunch out [and] told the person that I might not receive texts or calls.''}


Finally, 3 people who we tried to recruit as participants (not included in the 30 participants) declined to join the study
They felt that constant availability was expected at the work place and that without notifications they would not be able to maintain the expected level of availability. P09 -- who joined the study despite initial concerns -- said that she ``\emph{thought of saying `no' to take part}'' because she was ``\emph{worrying to miss calls from work}'' and she ``\emph{thought it would be horrible}.''

\begin{table*}[!ht]
\begin{center}
	\includegraphics[width=1.0\linewidth]{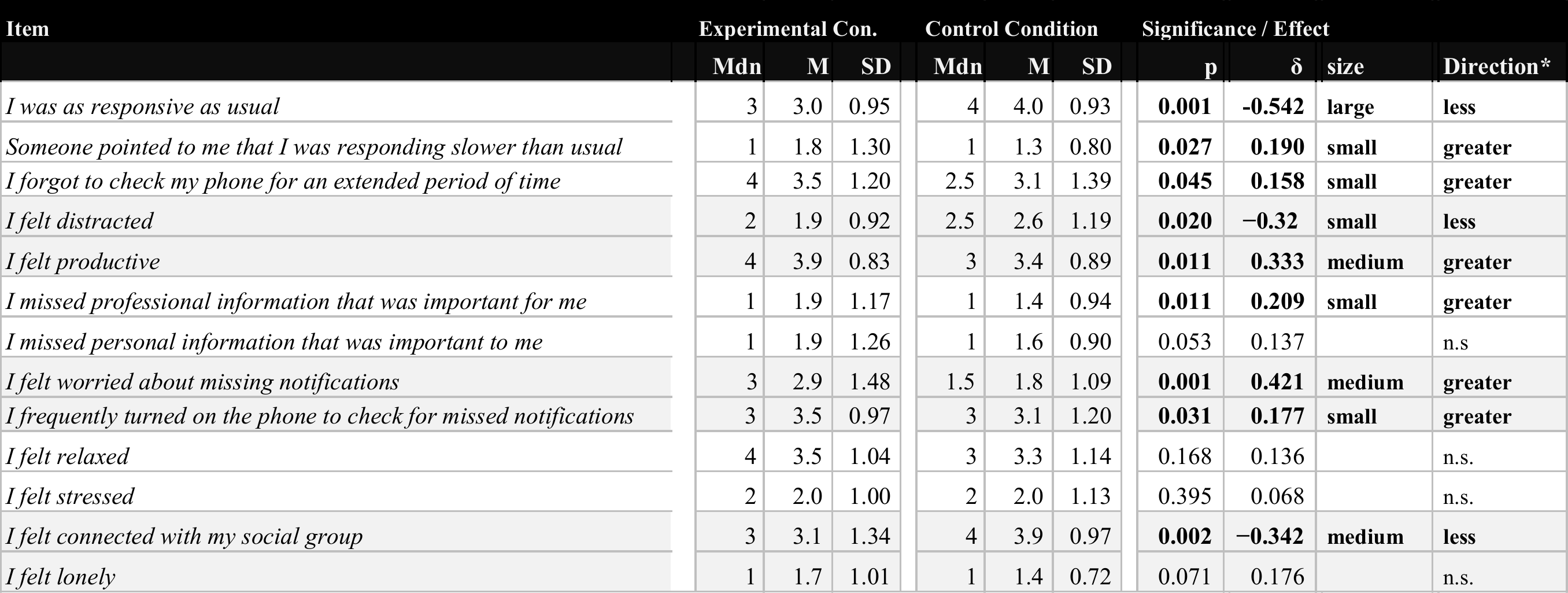}
	\caption{Statistical analysis of the responses to the questionnaires that were filled out after the days with and without notifications. Scores range from \textbf{1} (= disagree) to \textbf{5} (= agree). 
	The table lists median (Mdn), mean (M), and standard deviation (SD) of the responses for each condition.
	The right part of the table shows the results of the inferential tests and the magnitude of the effect (on the basis of Cliff's $\delta$).
	The direction* field indicates whether during the day without notifications (experimental condition), the agreement to the statement was significantly greater, less, or not significantly different.}
	\label{fig:ResultsTable}
\end{center}
\end{table*}

\section{Results}

All 30 participants successfully completed the \emph{Do Not Disturb} Challenge. During the interview, we confirmed that all 30 participants had complied with keeping notifications disabled or enabled, depending on the study condition. 

Table \ref{fig:ResultsTable} summarizes the quantitative results from the questionnaires that were issued on the day without notifications and the baseline day. For the descriptive statistics, we report median, mean, and standard deviation for each of the questionnaire items. 

When analyzing Likert scales, there is disagreement amongst scholars whether to use parametric or non-parametric tests. In this paper, we report the more conservative non-parametric statistics. Wilcoxon-Signed Rank test are used to test for significant differences and Cliff's \emph{delta} are used to estimate the effect size.
Please note that researchers have argued that for 5-point Likert scales, t-tests and Mann-Whitney-Wilcoxon tests have comparable power \cite{Winter2010}. As a test, we applied t-tests as well and found the same items to be significant. The use of parametric tests would have led to the same high-level conclusions.
In the following, we discuss the effects of the experimental manipulation on the survey responses.

\subsection{Drop in Engagement and Reduced Responsiveness}
The absence of notifications had a significant effect on how participants perceived their engagement with the mobile phone.
They agreed significantly less with the statement \statement{I was as responsive as usual} 
\result{$z = -3.269, p = 0.001$}.
The effect size ($\delta = -0.542$) suggests large practical significance.
At the same time, the agreement with the statement \statement{Someone pointed to me that I was responding slower than usual} was significantly higher 
\result{$z = -1.925, p = 0.027$}.
The effect size ($\delta = 0.158$) suggests low practical significance.
The participants' agreement with \statement{I forgot to check my phone for an extended period of time} was marginally higher when notifications were disabled
\result{$z = -1.698, p = 0.045$}.
The effect size ($\delta = 0.19$) suggests low practical significance.
As an example, P02 ``\emph{forgot my phone at work}'' because of not being reminded of the phone by notifications.
These effects indicate that subjective responsiveness and engagement with the phone decreased with the absence of notifications.

\subsection{Less Distraction and Higher Productivity}
The agreement to the statement \statement{I felt distracted} was significantly lower when notifications were disabled 
\result{$z = -2.054, p = 0.020$}.
The effect size ($\delta = -0.32$) suggests low practical significance.
In contrast, the agreement to the statement \statement{I felt productive} was significantly higher when notifications were disabled
\result{$z = -2.302, p = 0.011$}.
The effect size ($d = 0.333$) suggests medium practical significance.
P11 realized that ``\emph{after some time of frequently checking the phone for new notifications, I stopped checking, felt more productive.}''
P07 said that without notifications it was ``\emph{easier to concentrate, especially when working on the desktop.}''
These findings provide evidence that without notification alerts, the participants {felt less distracted and more productive}.

\subsection{Missing Information \& Violating Expectations}
During the day without notifications, participants were significantly more likely to agree with the statement
\statement{I missed professional information that was important for me} 
\result{$z = -2.289, p = 0.011, \delta = 0.209$, suggesting low practical significance}. Further, they were marginally more likely to agree with the statement
\statement{I missed personal information that was important for me} 
\result{$z = -1.615, p = 0.053, \delta = 0.137$, suggesting negligible practical significance}.
During the post-hoc interview, 8 participants reported to have missed important or urgent information. For example, P10 ``\emph{missed a WhatsApp group discussion, where my group decided to meet to sign a birthday postcard.''}
A friend of P02 ``\emph{was angry, saying that `we had a conversation and you forgot about it'.''} The girlfriend of P10 was only fine with delayed response time because ``\emph{she understood that this was part of a study}.''
These reports illustrate how the absence of notifications caused participants to miss information and violate expectations towards responsiveness.

\subsection{Worried about Missing Information}
The lack of notifications therefore created a new source of worry: the agreement with \statement{I felt worried about missing notifications} was significantly stronger 
\result{$z = -3.001, p = 0.001$} during the day without notifications.
The effect size ($\delta = 0.421$) suggests medium practical significance.
9 of the 30 participants reported that they were anxious to miss important or urgent information, such as,
\statement{I was waiting for a package and I was anxious to miss the call of the delivery service to notify me about the arrival} (P03) or simply
\statement{I felt like I was missing stuff} (P24).
Others had appointments and were afraid of missing messages. For example, P04 stated that  ``\emph{I was meeting with [a friend] for lunch, and I knew that I was going to receive something from her.}''

\subsection{More Frequent Checking}
This worry resulted into checking the phone more frequently. 
During the day without notifications, agreement to the statement \statement{I frequently turned on the phone to check for missed notifications}  was significantly higher \result{$z = -1.869, p = 0.031$}.
The effect size ($\delta = 0.177$) suggests low practical significance.
In the interview, 12 of the 30 participants reported to have checked their devices for new notifications more often that usual during the day without notification.
Comments related to this ranged from \statement{frequently checking my phone manually} (P07) to \statement{I even left the screen on not to miss [a friend's] notifications ... otherwise she would get angry} (P04).
Particularly extreme reactions were triggered when friends got angry: \statement{because of the reaction of my friend, who got angry because I forgot to respond, I was the whole afternoon with phone in my hand} (P12).
Some participants estimated the interval in which they began checking the phone. 
Interestingly, estimates named 30 minutes as interval lengths: \statement{I checked the phone every half hour} (P20), or \statement{Checked email ca. every 30 minutes} (P15).

\subsection{Stress \& Being Relaxed}
The anxiety induced by the absence of notifications did, however, not translate into a systematic increase in stress.
During the interview, 11 of the 30 participants reported from positive effects of not having notifications.
P09 said that ``\emph{Usually, I feel stressed, but in fact, today, I feel less stressed.}''
P03 found himself feeling ``\emph{more relaxed.}''
P22 concluded that \statement{It was amazing! I felt liberated!}
However, neither of the tests of the statements regarding stress (``\emph{I felt stressed}'' and ``\emph{I felt relaxed}'') revealed significant differences.
This might be explained by the finding that there are two opposing stress-inducing effects at work -- stress from the interruptions and stress from being anxious to miss important information or violate expectations --, which influenced participants to different extents.

\subsection{Feeling Less Connected With Others} 
Our study revealed a link between notifications and staying emotionally in touch with one's social group.
During the day without notifications, agreement to the statement \statement{I felt connected with my social group} was significantly lower 
\result{$z = -2.813, p = 0.002$}.
The effect size ($\delta = -0.342$) suggests medium practical significance.
The inverse statement \statement{I felt lonely} was, however, only marginally significant 
\result{$z = -1.471, p = 0.071$}. The effect size ($\delta = 0.176$) would have suggested a small practical significance.
These results contrast that -- while work-wise, disabling notifications helped to be more focussed and productive -- socially, they negatively affect the feeling of being in touch with one's social group.

\vfill\eject
\section{Post-Study Reflections}

The participants' post-study reflections to having notifications disabled varied greatly. They ranged from 
very positive responses, such as
\statement{It was amazing! I felt liberated!} (P22)
over neutral responses
\statement{It was not a big deal, since I am usually not checking notifications and people know that I am not responsive} (P25)
to very negative responses
\statement{I was paranoid and I even left the screen on not to miss a friends notification} (P04).
The strong reactions on both ends emphasize the magnitude of the effect that notifications have on some people's lives.

\subsection{Manage Notifications More Consciously}
9 of the 30 participants reported that thanks to their participation in the \emph{Do Not Disturb Challenge}, they would {manage notifications more consciously} in the future. 
For example, P09 
\statement{got aware how much [WhatsApp-group notifications] are stressing me}.
P20 was \statement{considering to only keep notifications for the important things, so people can better reach me} and P26 had come to the conclusion that \statement{The important apps are Messenger, Hangout and WhatsApp. The rest does not require notifications}. P14 added \statement{SMS} to the list.
This shows that our participants were becoming aware that not all notifications are important, and that for them the most important source of notifications are messaging apps.

\subsection{Using {Do Not Disturb} Mode in the Future}
13 of the 30 participants said that they would use \emph{Do Not Disturb} or similar notification-suppression modes in the future.
11 of them planned to disable notifications during specific times or activities, such as: 
\statement{put the phone in \emph{Do Not Disturb} Mode when I study} (P18),
\statement{for reading papers, want to concentrate} (P07), or
 \statement{when I need to really get things done, I need to turn notifications off} (P24).
2 participants decided to keep \emph{Do Not Disturb} permanently enabled.

\vfill\eject
\subsection{Two Years Later - Did Participants Follow Through?}
In April 2017, two years after the study had been conducted, we contacted the 22 participants who intended to disable notifications selectively or use Do Not Disturb in the future. We reminded them of the intentions that they had expressed during the interview, and asked them whether they had followed through with these intentions. 
13 of those 22 participants (59.1\%) followed through with their plans. For example, P14, who planned to only keep notifications enabled for important applications responded: 
\statement{I have followed through with my original plan of keeping only important messages from SMS, none from Facebook or other social media.}
4 (18.2\%) followed through partially. For example, P11, who planned to disabled Skype notification balloons, responded \statement{I disabled for Skype personal, but re-enabled for professional.}
3 (13.6\%) did not follow through at all. For example, P9 who planned to disable WhatsApp group notifications, as she got aware how much they are stressing, responded: \statement{Unfortunately I'm not following the plan and I haven't disabled my Whatsapp group notifications. Probably I got used to having stress around ;)}
2 of the 22 (9.1\%) participants did not respond to our inquiry.

\section{Discussion}

The \emph{Do Not Disturb} Challenge revealed strong and polarized reactions to the absence of notifications.
For some participants, being without notifications was a positive experience: being more relaxed, less stressed, and more productive at work. 
For others, fear of missing out and violating others' expectations turned it into a negative experience.

\subsection{Notifications Drive Phone Use and Distract}
The absence of notifications had a significant effect on the participants' subjective responsiveness. 
During the day without notifications, participants were significantly more likely to feel less responsive than usual, and it was more often pointed out to them that they responded slower than usual.
Further, without notifications, participants reported to have been more likely to forget checking the phone for extended periods of time.
This evidence corroborates previous findings by Lee \emph{et al.} \cite{Lee2014} that in mobile phone usage is often triggered by notifications. It also corroborates previous work that notifications cause people to interrupt current activities to timely triage the notification \cite{Iqbal2010,Mark2012,Pielot2014notif}.
In contrast to Kushlev \emph{et al.} \cite{Kushlev2016}, we did not ask to our participants to keep their devices ``out of sight, out of mind''. Thus, our study design was less likely to limit self-interruptions.

During the day without notifications, our participants reported to feel significantly more productive and less distracted.
This confirms a long history of findings that notifications interrupt \cite{Borst2015,Iqbal2010,Pielot2014notif} and have negative effects on task performance \cite{Adamczyk2004,Borst2015,Cutrell2001,Czerwinski2004,Iqbal2008,Iqbal2010,Kushlev2016,Leiva2012,Mark2012,Okoshi2015,Wajcman2011,Stothart2015}.
This strengthens the need for research about delivering notifications at opportune moments \cite{Iqbal2008,Mehrotra2015,Okoshi2015}. 
However, some participants of our study also expressed that they did not feel interrupted by notifications, which might be explained by the finding the interruptions are perceived differently, depending on the nature of the concurrent activity \cite{Mark2015}, and that productivity impairments can be largely explained by the inattention introduced by notifications \cite{Kushlev2016}. 

In summary, the absence of notifications made our participants engage less often with the phone, decreased perceived distractions, and increased self-perceived productivity.

\subsection{Notifications are Essential to Meet Social Expectations}
On the one hand, the absence of notifications had clear positive effects. On the other hand, the absence of notifications became a new source anxiety and significantly increased the worry to miss information.

In one-third of the interviews, social expectations came up as the number one reason. 
As reported in previous work \cite{Birnholtz2012,Church2013,Pielot2014notif}, we found that the majority of notifications originates from communication applications, where not responding timely can be perceived as an offense to the sender. 
In the pre-study questionnaire, 80\% of the participants agreed with the statement that they are expected to respond timely. In the post-study interview, our participants reported numerous anecdotes, in which missed messages had lead to conflicts with friends and partners.

In another third of the interviews, participants emphasized the expectations of the work place to respond timely as major issue.
In the email-notification-deprivation study by Iqbal and Horvitz from 2010, where participants re-enabled work email notifications, many of them said that they did so for the \emph{awareness} that notifications provided rather than because they felt that they \emph{had} to re-enable them \cite{Iqbal2010}.
However, more recently (2014), Mazmanian and Erickson \cite{Mazmanian2014} argued that constant availability has become part of the offer that companies make to their customers. And in fact, many information workers allow work emails to cross the boundaries between work and personal life \cite{Cecchinato2015,Chen2014}. This development may explain why 3 people declined participation in the study: they felt that without notifications they would not be able to comply with the expectations of the work place.

The worry to miss important information or violate social expectations was so serious that 40\% of the participants reported to react to this worry by frequently checking the phone when they were expecting important messages. 
Without notifications, they felt no longer able to maintain the expected level of availability -- which had been enabled by notifications in the first place. 
This provides evidence that disabling notifications -- even though it reduces the number of unwanted and stressful distractions -- puts many people into a situation that for some of them has worse impact on their affective state than keeping notifications enabled.

However, we must note that not all participants were subject to this new source of anxiety. One salient factor was that participants who had no issue with being without notification was that others already knew that they usually would not respond timely to messages. We hypothesize that managing the expectations of frequent communication partners with regards to response times may be key to reduce notification- and texting-induced technostress.

\subsection{Notifications Connect}
When notifications were enabled, there was higher agreement to the statement \statement{I feel connected with my social group}. This indicates that without notifications, our participants felt less connected with others.
From the interviews, we learned that for our participants, notifications were largely related to personal communication services.
Our findings confirm previous work \cite{Pielot2014notif,Sahami2014} that notifications from messaging applications are deemed the most important.
The participants who planned to selectively disable notifications as a result of the study, were frequently stating that notifications from these type of apps would stay enabled, \emph{e.g.}, \statement{I keep WhatsApp} (P11) or \statement{I am considering to only keep notifications for the important things, so people can better reach me} (P20).

Given the negative effects that notifications can have, we may be tempted to demonize them. However, these findings remind us that \finding{notifications} are also something positive, as they \finding{had the effect of making the participants feel more connected with the people they care about}.

\subsection{Notification Overload}
The most novel insight, which has never been reported in prior work thus far, is that more than two-third of our participants reported to planned changes to the way they manage notifications. In contrast, in the study on disabling email notifications by Iqbal and Horvitz \cite{Iqbal2010}, all participants re-enabled notifications after the study.

{One third} of the participants reported to be \finding{selectively disabling} notifications after having participated in the \emph{Do Not Disturb} Challenge. Some disabled WhatsApp group notifications, others disabled all notifications from all apps except messengers. This confirms previous findings \cite{Mehrotra2015,Sahami2014} that notifications from communication services are more important than others.

{Almost half} of the participants stated that they would use notification-suppressing settings, such as \emph{Do Not Disturb}, to disable all notifications in the future. The most-frequently named intention was to disable notifications during work time, in order to improve concentration and productivity.
However, two participants said that they would be keeping notifications disabled around the clock in the future.

Two years after the study, about 59.1\% of the 22 participants who expressed such intentions are still following through with them. 77.3\% of these participants are still following through partially. Since mobile phone users rarely change notification settings \cite{Chang2015,Westermann2015}, this emphasizes the magnitude of the effect that the \emph{Do Not Disturb} study had on the participants.
The fact that more than half of the participants reduced the number of notifications that they are exposed to on a daily basis is a warning sign that \finding{our participants were realizing a sense notification overload.}  

Today, we are still living in the ``wild-west land-grab phase'' of notifications: more and more platforms (OSes, browsers, ...) introduce push-notification channels. An increasing number of apps and services is subjecting its users to notifications.
Our study highlights one potential outcome of this development: if apps and services do not treat people's attention with care and subject them to an ever increasing number of notifications, they may suffer the Tragedy of the Commons \cite{Hardin1968}. More and more people may follow the example of our participants, and consider the use of more drastic measures to take back control, \emph{e.g.}, by disabling notifications for specific applications or disabling all notifications during specific phases. In the long run, this may significantly limit the usefulness of notifications to drive engagement, to connect people, and to deliver proactive recommendations. This is a clear call for using notifications responsibly, \emph{i.e}, to ensure good timing and relevance of notifications.

\subsection{Limitations}
Our participants were a sample of 30 white-collar workers. The results may not generalize to other segments of the population.
In particular, the findings may not apply for people who cannot use computer-mediated communication tools at work, who cannot check notifications for extended periods of time, or who are not too occupied by their daily activities and therefore welcome distractions in general.
The study relies on self-reported data. Thus, findings are based on the participants' self perception, which can suffer from biases. 
Further, single-item scales give holistic insights related to a feeling (e.g. being stressed), but they cannot necessarily distinguish the exact underlying factors (e.g. the exact type of stress).
Initially, we tried to recruit participants for a one-week period without notifications. When too many people declined to participate, because they felt that this period was too long, we limited it to 24 hours to avoid self-selection bias. As a consequence, participants had very little time to accustom themselves to the lack of notifications. We assume that over time that magnitude of the observed effects may change.

\section{Conclusions}

We present an experimental study to investigate the effect that notifications across all devices and services have on its users. 
In order to isolate notifications as \emph{cause}, we asked 30 people to disable notifications for a day, and compared self-reported behaviors and emotions to a baseline day.
The data we collected shows strong and polarized reactions to being without notifications, revealing a critical contrast:

\begin{itemize}\compresslist
	\item Notifications negatively impacted focused work, as participants reported to feel significantly less distracted and more productive without them.
	\item At the same time, disabling notifications also had significant negative effects: it made participants more worried to miss important information, not being responsive enough, and feeling less connected with their social network.
	\item In contrast to a previous deprivation study, where all participants re-enabled work email notifications after the study, about one-third of our participants expressed the intention to disable some sources of notifications, and about half of our participants expressed the intention to use Do Not Disturb (and equivalent settings) more often in the future. Two years later, 60\% of these participants are still following through with their intentions. Another 18\% have changed their notification-related behavior.
\end{itemize}

Our findings show that cultural practices around notifications have locked people in a dilemma: on the one hand, notifications have become integral to the tools that connect us with others, and they are needed to keep up with people's expectations. On the other hand, our participants became aware of the negative effects that notifications have on them and some started to devise coping strategies. Notifications as a channel to engage with people may be threatened if this dilemma is not addressed.



\bibliographystyle{SIGCHI-Reference-Format}

\begin{thebibliography}{00}


\ifx \showCODEN    \undefined \def \showCODEN     #1{\unskip}     \fi
\ifx \showDOI      \undefined \def \showDOI       #1{{\tt DOI:}\penalty0{#1}\ }
  \fi
\ifx \showISBNx    \undefined \def \showISBNx     #1{\unskip}     \fi
\ifx \showISBNxiii \undefined \def \showISBNxiii  #1{\unskip}     \fi
\ifx \showISSN     \undefined \def \showISSN      #1{\unskip}     \fi
\ifx \showLCCN     \undefined \def \showLCCN      #1{\unskip}     \fi
\ifx \shownote     \undefined \def \shownote      #1{#1}          \fi
\ifx \showarticletitle \undefined \def \showarticletitle #1{#1}   \fi
\ifx \showURL      \undefined \def \showURL       #1{#1}          \fi

\bibitem{Adamczyk2004}
{Piotr~D. Adamczyk} {and} {Brian~P. Bailey}. 2004.
\newblock \showarticletitle{If Not Now, when?: The Effects of Interruption at
  Different Moments Within Task Execution}. In {\em Proceedings of the SIGCHI
  Conference on Human Factors in Computing Systems} {\em (CHI '04)}. ACM, New
  York, NY, USA, 271--278.
\newblock
\showISBNx{1-58113-702-8}
\showDOI{%
\url{http://dx.doi.org/10.1145/985692.985727}}


\bibitem{Battestini2010}
{Agathe Battestini}, {Vidya Setlur}, {and} {Timothy Sohn}. 2010.
\newblock \showarticletitle{A Large Scale Study of Text-messaging Use}. In {\em
  Proceedings of the 12th International Conference on Human Computer
  Interaction with Mobile Devices and Services} {\em (MobileHCI '10)}. ACM, New
  York, NY, USA, 229--238.
\newblock
\showISBNx{978-1-60558-835-3}
\showDOI{%
\url{http://dx.doi.org/10.1145/1851600.1851638}}


\bibitem{Birnholtz2012}
{Jeremy Birnholtz}, {Jeff Hancock}, {Madeline Smith}, {and} {Lindsay Reynolds}.
  2012.
\newblock \showarticletitle{Understanding unavailability in a world of constant
  connection}.
\newblock {\em interactions\/} {19}, 5 (2012), 32--35.
\newblock
\showISSN{1072-5520}
\showDOI{%
\url{http://dx.doi.org/10.1145/2334184.2334193}}


\bibitem{Borst2015}
{Jelmer~P. Borst}, {Niels~A. Taatgen}, {and} {Hedderik van Rijn}. 2015.
\newblock \showarticletitle{What Makes Interruptions Disruptive?: A
  Process-Model Account of the Effects of the Problem State Bottleneck on Task
  Interruption and Resumption}. In {\em Proceedings of the 33rd Annual ACM
  Conference on Human Factors in Computing Systems} {\em (CHI '15)}. ACM, New
  York, NY, USA, 2971--2980.
\newblock
\showISBNx{978-1-4503-3145-6}
\showDOI{%
\url{http://dx.doi.org/10.1145/2702123.2702156}}


\bibitem{Braun2006}
{Virginia Braun} {and} {Victoria Clarke}. 2006.
\newblock \showarticletitle{Using thematic analysis in psychology}.
\newblock {\em Qualitative Research in Psychology\/} {3}, 2 (2006), 77--101.
\newblock
\showDOI{%
\url{http://dx.doi.org/10.1191/1478088706qp063oa}}


\bibitem{Cecchinato2015}
{Marta~E. Cecchinato}, {Anna~L. Cox}, {and} {Jon Bird}. 2015.
\newblock \showarticletitle{Working 9-5?: Professional Differences in Email and
  Boundary Management Practices}. In {\em Proceedings of the 33rd Annual ACM
  Conference on Human Factors in Computing Systems} {\em (CHI '15)}. ACM, New
  York, NY, USA, 3989--3998.
\newblock
\showISBNx{978-1-4503-3145-6}
\showDOI{%
\url{http://dx.doi.org/10.1145/2702123.2702537}}


\bibitem{Chang2015}
{Yung-Ju Chang} {and} {John~C. Tang}. 2015.
\newblock \showarticletitle{Investigating Mobile Users' Ringer Mode Usage and
  Attentiveness and Responsiveness to Communication}. In {\em Proceedings of
  the 17th International Conference on Human-Computer Interaction with Mobile
  Devices and Services} {\em (MobileHCI '15)}. ACM, New York, NY, USA, 6--15.
\newblock
\showISBNx{978-1-4503-3652-9}
\showDOI{%
\url{http://dx.doi.org/10.1145/2785830.2785852}}


\bibitem{Chen2014}
{Adela Chen} {and} {Elena Karahanna}. 2014.
\newblock \showarticletitle{Boundaryless Technology: Understanding the Effects
  of Technology-Mediated Interruptions across the Boundaries between Work and
  Personal Life}.
\newblock {\em AIS Transactions on Human-Computer Interaction\/} {6}, 2 (2014),
  16--36.
\newblock


\bibitem{Church2013}
{Karen Church} {and} {Rodrigo de Oliveira}. 2013.
\newblock \showarticletitle{What's Up with Whatsapp?: Comparing Mobile Instant
  Messaging Behaviors with Traditional SMS}. In {\em Proceedings of the 15th
  International Conference on Human-computer Interaction with Mobile Devices
  and Services} {\em (MobileHCI '13)}. ACM, New York, NY, USA, 352--361.
\newblock
\showISBNx{978-1-4503-2273-7}
\showDOI{%
\url{http://dx.doi.org/10.1145/2493190.2493225}}


\bibitem{Cutrell2001}
{Ed Cutrell}, {Mary Czerwinski}, {and} {Eric Horvitz}. 2001.
\newblock \showarticletitle{Notification, Disruption, and Memory: Effects of
  Messaging Interruptions on Memory and Performance}. In {\em Proc. INTERACT
  '01}. IOS Press.
\newblock


\bibitem{Czerwinski2004}
{Mary Czerwinski}, {Eric Horvitz}, {and} {Susan Wilhite}. 2004.
\newblock \showarticletitle{A Diary Study of Task Switching and Interruptions}.
  In {\em Proceedings of the SIGCHI Conference on Human Factors in Computing
  Systems} {\em (CHI '04)}. ACM, New York, NY, USA, 175--182.
\newblock
\showISBNx{1-58113-702-8}
\showDOI{%
\url{http://dx.doi.org/10.1145/985692.985715}}


\bibitem{Winter2010}
{Joost C.~F. de Winter} {and} {Dimitra Dodou}. 2010.
\newblock \showarticletitle{Five-Point Likert Items: t test versus
  Mann-Whitney-Wilcoxon}.
\newblock {\em Practical Assessment, Research \& Evaluation\/} {15}, 11
  (October 2010), 1--12.
\newblock
\showURL{%
\url{http://doi.acm.org/10.1145/985692.985727}}


\bibitem{Dingler2015}
{Tilman Dingler} {and} {Martin Pielot}. 2015.
\newblock \showarticletitle{I'll Be There for You: Quantifying Attentiveness
  Towards Mobile Messaging}. In {\em Proceedings of the 17th International
  Conference on Human-Computer Interaction with Mobile Devices and Services}
  {\em (MobileHCI '15)}. ACM, New York, NY, USA, 1--5.
\newblock
\showISBNx{978-1-4503-3652-9}
\showDOI{%
\url{http://dx.doi.org/10.1145/2785830.2785840}}


\bibitem{Gallud2015}
{Jose~A. Gallud} {and} {Ricardo Tesoriero}. 2015.
\newblock \showarticletitle{Smartphone Notifications: A Study on the Sound to
  Soundless Tendency}. In {\em Proceedings of the 17th International Conference
  on Human-Computer Interaction with Mobile Devices and Services Adjunct} {\em
  (MobileHCI '15)}. ACM, New York, NY, USA, 819--824.
\newblock
\showISBNx{978-1-4503-3653-6}
\showDOI{%
\url{http://dx.doi.org/10.1145/2786567.2793706}}


\bibitem{Hardin1968}
{Garrett Hardin}. 1968.
\newblock \showarticletitle{The Tragedy of the Commons}.
\newblock {\em Science\/} {162}, 3859 (1968), 1243ñ1248.
\newblock
\showDOI{%
\url{http://dx.doi.org/doi:10.1126/science.162.3859.1243}}


\bibitem{Iqbal2008}
{Shamsi~T. Iqbal} {and} {Brian~P. Bailey}. 2008.
\newblock \showarticletitle{Effects of Intelligent Notification Management on
  Users and Their Tasks}. In {\em Proceedings of the SIGCHI Conference on Human
  Factors in Computing Systems} {\em (CHI '08)}. ACM, New York, NY, USA,
  93--102.
\newblock
\showISBNx{978-1-60558-011-1}
\showDOI{%
\url{http://dx.doi.org/10.1145/1357054.1357070}}


\bibitem{Iqbal2010a}
{Shamsi~T. Iqbal} {and} {Brian~P. Bailey}. 2010.
\newblock \showarticletitle{Oasis: A framework for linking notification
  delivery to the perceptual structure of goal-directed tasks}.
\newblock {\em ACM Trans. Comput.-Hum. Interact.\/} {17}, 4, Article 15 (Dec
  2010), 28 pages.
\newblock
\showISSN{1073-0516}
\showDOI{%
\url{http://dx.doi.org/10.1145/1879831.1879833}}


\bibitem{Iqbal2010}
{Shamsi~T. Iqbal} {and} {Eric Horvitz}. 2010.
\newblock \showarticletitle{Notifications and Awareness: A Field Study of Alert
  Usage and Preferences}. In {\em Proceedings of the 2010 ACM Conference on
  Computer Supported Cooperative Work} {\em (CSCW '10)}. ACM, New York, NY,
  USA, 27--30.
\newblock
\showISBNx{978-1-60558-795-0}
\showDOI{%
\url{http://dx.doi.org/10.1145/1718918.1718926}}


\bibitem{Jin2009}
{Jing Jin} {and} {Laura~A. Dabbish}. 2009.
\newblock \showarticletitle{Self-interruption on the Computer: A Typology of
  Discretionary Task Interleaving}. In {\em Proceedings of the SIGCHI
  Conference on Human Factors in Computing Systems} {\em (CHI '09)}. ACM, New
  York, NY, USA, 1799--1808.
\newblock
\showISBNx{978-1-60558-246-7}
\showDOI{%
\url{http://dx.doi.org/10.1145/1518701.1518979}}


\bibitem{Kushlev2014}
{Kostadin Kushlev} {and} {Elizabeth~W. Dunn}. 2014.
\newblock \showarticletitle{Checking email less frequently reduces stress}.
\newblock {\em Computers in Human Behavior\/}  {43} (2014), 220--228.
\newblock


\bibitem{Kushlev2016}
{Kostadin Kushlev}, {Jason Proulx}, {and} {Elizabeth~W. Dunn}. 2016.
\newblock \showarticletitle{"Silence Your Phones": Smartphone Notifications
  Increase Inattention and Hyperactivity Symptoms}. In {\em Proceedings of the
  2016 CHI Conference on Human Factors in Computing Systems} {\em (CHI '16)}.
  ACM, New York, NY, USA, 1011--1020.
\newblock
\showISBNx{978-1-4503-3362-7}
\showDOI{%
\url{http://dx.doi.org/10.1145/2858036.2858359}}


\bibitem{Lee2014}
{Uichin Lee}, {Joonwon Lee}, {Minsam Ko}, {Changhun Lee}, {Yuhwan Kim}, {Subin
  Yang}, {Koji Yatani}, {Gahgene Gweon}, {Kyong-Mee Chung}, {and} {Junehwa
  Song}. 2014.
\newblock \showarticletitle{Hooked on Smartphones: An Exploratory Study on
  Smartphone Overuse Among College Students}. In {\em Proceedings of the SIGCHI
  Conference on Human Factors in Computing Systems} {\em (CHI '14)}. ACM, New
  York, NY, USA, 2327--2336.
\newblock
\showISBNx{978-1-4503-2473-1}
\showDOI{%
\url{http://dx.doi.org/10.1145/2556288.2557366}}


\bibitem{Leiva2012}
{Luis Leiva}, {Matthias B\"{o}hmer}, {Sven Gehring}, {and} {Antonio
  Kr\"{u}ger}. 2012.
\newblock \showarticletitle{Back to the App: The Costs of Mobile Application
  Interruptions}. In {\em Proceedings of the 14th International Conference on
  Human-computer Interaction with Mobile Devices and Services} {\em (MobileHCI
  '12)}. ACM, New York, NY, USA, 291--294.
\newblock
\showISBNx{978-1-4503-1105-2}
\showDOI{%
\url{http://dx.doi.org/10.1145/2371574.2371617}}


\bibitem{Tovar2015}
{Hugo Lopez-Tovar}, {Andreas Charalambous}, {and} {John Dowell}. 2015.
\newblock \showarticletitle{Managing Smartphone Interruptions Through Adaptive
  Modes and Modulation of Notifications}. In {\em Proceedings of the 20th
  International Conference on Intelligent User Interfaces} {\em (IUI '15)}.
  ACM, New York, NY, USA, 296--299.
\newblock
\showISBNx{978-1-4503-3306-1}
\showDOI{%
\url{http://dx.doi.org/10.1145/2678025.2701390}}


\bibitem{Mark2008}
{Gloria Mark}, {Daniela Gudith}, {and} {Ulrich Klocke}. 2008.
\newblock \showarticletitle{The Cost of Interrupted Work: More Speed and
  Stress}. In {\em Proceedings of the SIGCHI Conference on Human Factors in
  Computing Systems} {\em (CHI '08)}. ACM, New York, NY, USA, 107--110.
\newblock
\showISBNx{978-1-60558-011-1}
\showDOI{%
\url{http://dx.doi.org/10.1145/1357054.1357072}}


\bibitem{Mark2015}
{Gloria Mark}, {Shamsi Iqbal}, {Mary Czerwinski}, {and} {Paul Johns}. 2015.
\newblock \showarticletitle{Focused, Aroused, but So Distractible: Temporal
  Perspectives on Multitasking and Communications}. In {\em Proceedings of the
  18th ACM Conference on Computer Supported Cooperative Work \&\#38; Social
  Computing} {\em (CSCW '15)}. ACM, New York, NY, USA, 903--916.
\newblock
\showISBNx{978-1-4503-2922-4}
\showDOI{%
\url{http://dx.doi.org/10.1145/2675133.2675221}}


\bibitem{Mark2012}
{Gloria Mark}, {Stephen Voida}, {and} {Armand Cardello}. 2012.
\newblock \showarticletitle{"A Pace Not Dictated by Electrons": An Empirical
  Study of Work Without Email}. In {\em Proceedings of the SIGCHI Conference on
  Human Factors in Computing Systems} {\em (CHI '12)}. ACM, New York, NY, USA,
  555--564.
\newblock
\showISBNx{978-1-4503-1015-4}
\showDOI{%
\url{http://dx.doi.org/10.1145/2207676.2207754}}


\bibitem{Mazmanian2014}
{Melissa Mazmanian} {and} {Ingrid Erickson}. 2014.
\newblock \showarticletitle{The Product of Availability: Understanding the
  Economic Underpinnings of Constant Connectivity}. In {\em Proceedings of the
  SIGCHI Conference on Human Factors in Computing Systems} {\em (CHI '14)}.
  ACM, New York, NY, USA, 763--772.
\newblock
\showISBNx{978-1-4503-2473-1}
\showDOI{%
\url{http://dx.doi.org/10.1145/2556288.2557381}}


\bibitem{Mehrotra2015}
{Abhinav Mehrotra}, {Mirco Musolesi}, {Robert Hendley}, {and} {Veljko Pejovic}.
  2015.
\newblock \showarticletitle{Designing Content-driven Intelligent Notification
  Mechanisms for Mobile Applications}. In {\em Proceedings of the 2015 ACM
  International Joint Conference on Pervasive and Ubiquitous Computing} {\em
  (UbiComp '15)}. ACM, New York, NY, USA, 813--824.
\newblock
\showISBNx{978-1-4503-3574-4}
\showDOI{%
\url{http://dx.doi.org/10.1145/2750858.2807544}}


\bibitem{Okoshi2015}
{Tadashi Okoshi}, {Julian Ramos}, {Hiroki Nozaki}, {Jin Nakazawa}, {Anind~K.
  Dey}, {and} {Hideyuki Tokuda}. 2015.
\newblock \showarticletitle{Reducing Users' Perceived Mental Effort Due to
  Interruptive Notifications in Multi-device Mobile Environments}. In {\em
  Proceedings of the 2015 ACM International Joint Conference on Pervasive and
  Ubiquitous Computing} {\em (UbiComp '15)}. ACM, New York, NY, USA, 475--486.
\newblock
\showISBNx{978-1-4503-3574-4}
\showDOI{%
\url{http://dx.doi.org/10.1145/2750858.2807517}}


\bibitem{Oulasvirta2012}
{Antti Oulasvirta}, {Tye Rattenbury}, {Lingyi Ma}, {and} {Eeva Raita}. 2012.
\newblock \showarticletitle{Habits make smartphone use more pervasive}.
\newblock {\em Personal and Ubiquitous Computing\/} {16}, 1 (2012), 105--114.
\newblock


\bibitem{Patterson2008}
{Donald~J. Patterson}, {Christopher Baker}, {Xianghua Ding}, {Samuel~J.
  Kaufman}, {Kah Liu}, {and} {Andrew Zaldivar}. 2008.
\newblock \showarticletitle{Online Everywhere: Evolving Mobile Instant
  Messaging Practices}. In {\em Proceedings of the 10th International
  Conference on Ubiquitous Computing} {\em (UbiComp '08)}. ACM, New York, NY,
  USA, 64--73.
\newblock
\showISBNx{978-1-60558-136-1}
\showDOI{%
\url{http://dx.doi.org/10.1145/1409635.1409645}}


\bibitem{Pielot2014notif}
{Martin Pielot}, {Karen Church}, {and} {Rodrigo de Oliveira}. 2014a.
\newblock \showarticletitle{An In-situ Study of Mobile Phone Notifications}. In
  {\em Proceedings of the 16th International Conference on Human-computer
  Interaction with Mobile Devices \&\#38; Services} {\em (MobileHCI '14)}. ACM,
  New York, NY, USA, 233--242.
\newblock
\showISBNx{978-1-4503-3004-6}
\showDOI{%
\url{http://dx.doi.org/10.1145/2628363.2628364}}


\bibitem{Pielot2014attpred}
{Martin Pielot}, {Rodrigo de Oliveira}, {Haewoon Kwak}, {and} {Nuria Oliver}.
  2014b.
\newblock \showarticletitle{Didn't You See My Message?: Predicting
  Attentiveness to Mobile Instant Messages}. In {\em Proceedings of the 32Nd
  Annual ACM Conference on Human Factors in Computing Systems} {\em (CHI '14)}.
  ACM, New York, NY, USA, 3319--3328.
\newblock
\showISBNx{978-1-4503-2473-1}
\showDOI{%
\url{http://dx.doi.org/10.1145/2556288.2556973}}


\bibitem{Pielot2015dnd}
{Martin Pielot} {and} {Luz Rello}. 2015.
\newblock \showarticletitle{The Do Not Disturb Challenge: A Day Without
  Notifications}. In {\em Proceedings of the 33rd Annual ACM Conference
  Extended Abstracts on Human Factors in Computing Systems} {\em (CHI EA '15)}.
  ACM, New York, NY, USA, 1761--1766.
\newblock
\showISBNx{978-1-4503-3146-3}
\showDOI{%
\url{http://dx.doi.org/10.1145/2702613.2732704}}


\bibitem{Rennecker2005}
{Julie Rennecker} {and} {Lindsey Godwin}. 2005.
\newblock \showarticletitle{Delays and Interruptions: A Self-perpetuating
  Paradox of Communication Technology Use}.
\newblock {\em Inf. Organ.\/} {15}, 3 (July 2005), 247--266.
\newblock
\showISSN{1471-7727}
\showDOI{%
\url{http://dx.doi.org/10.1016/j.infoandorg.2005.02.004}}


\bibitem{Sahami2014}
{Alireza Sahami~Shirazi}, {Niels Henze}, {Tilman Dingler}, {Martin Pielot},
  {Dominik Weber}, {and} {Albrecht Schmidt}. 2014.
\newblock \showarticletitle{Large-scale Assessment of Mobile Notifications}. In
  {\em Proceedings of the SIGCHI Conference on Human Factors in Computing
  Systems} {\em (CHI '14)}. ACM, New York, NY, USA, 3055--3064.
\newblock
\showISBNx{978-1-4503-2473-1}
\showDOI{%
\url{http://dx.doi.org/10.1145/2556288.2557189}}


\bibitem{Stothart2015}
{Cary Stothart}, {Ainsley Mitchum}, {and} {Courtney Yehnert}. 2015.
\newblock \showarticletitle{The attentional cost of receiving a cell phone
  notification.}
\newblock {\em Journal of experimental psychology: human perception and
  performance\/} {41}, 4 (2015), 893.
\newblock
\showURL{%
\url{http://dx.doi.org/10.1037/xhp0000100}}


\bibitem{Wajcman2011}
{Judy Wajcman} {and} {Emily Rose}. 2011.
\newblock \showarticletitle{Constant Connectivity: Rethinking Interruptions at
  Work}.
\newblock {\em Organization Studies\/} {32}, 7 (2011), 941--961.
\newblock


\bibitem{Wanous1997}
{JP Wanous}, {AE Reichers}, {and} {MJ Hudy}. 1997.
\newblock \showarticletitle{Overall Job Satisfaction: How Good Are Single-Item
  Measures?}
\newblock {\em J Appl Psychol\/} {82}, 2 (April 1997), 247--252.
\newblock


\bibitem{Westermann2015}
{Tilo Westermann}, {Sebastian M\"{o}ller}, {and} {Ina Wechsung}. 2015.
\newblock \showarticletitle{Assessing the Relationship between Technical
  Affinity, Stress and Notifications on Smartphones}. In {\em Proceedings of
  the 17th International Conference on Human-Computer Interaction with Mobile
  Devices and Services Adjunct} {\em (MobileHCI '15)}. ACM, New York, NY, USA,
  652--659.
\newblock
\showISBNx{978-1-4503-3653-6}
\showDOI{%
\url{http://dx.doi.org/10.1145/2786567.2793684}}


\bibitem{Youngblut1993}
{Joanne~M. Youngblut} {and} {Gail~R. Casper}. 1993.
\newblock \showarticletitle{Focus on Psychometrics Single-item Indicators in
  Nursing Research}.
\newblock {\em Research in Nursing and Health\/} {16}, 6 (December 1993),
  459--465.
\newblock


\end{thebibliography}


\balance

\end{document}